\documentclass[conference,10pt]{IEEEtran}
\usepackage{epsfig,rotating,setspace,latexsym,amsmath,epsf,amssymb,amsfonts,bm,theorem,subfigure,epstopdf}
\usepackage{cite,authblk}
\usepackage{bbm}
\usepackage[ruled,vlined]{algorithm2e}

\usepackage{color}

\DeclareMathOperator*{\argminA}{arg\,min}

\SetArgSty{textnormal}

\IEEEoverridecommandlockouts
\allowdisplaybreaks

\begin{document}
\bstctlcite{IEEEexample:BSTcontrol} 

\title{Channel Effects on Surrogate Models of Adversarial Attacks against Wireless Signal Classifiers}
	
\author[1]{Brian Kim}
\author[2]{Yalin E. Sagduyu}
\author[2]{Tugba Erpek}
\author[2]{Kemal Davaslioglu}
\author[1]{Sennur Ulukus}

	
\affil[1]{\normalsize Department of Electrical and Computer Engineering, University of Maryland, College Park, MD 20742, USA}
\affil[2]{\normalsize Intelligent Automation, Inc., Rockville, MD 20855, USA  \thanks{This effort is supported by the U.S. Army Research Office under contract W911NF-17-C-0090. The content of the information does not necessarily reflect the position or the policy of the U.S. Government, and no official endorsement should be inferred.}}
	
\maketitle

\begin{abstract}

We consider a wireless communication system that consists of a background emitter, a transmitter, and an adversary. The transmitter is equipped with a deep neural network (DNN) classifier for detecting the ongoing transmissions from the background emitter and transmits a signal if the spectrum is idle. Concurrently, the adversary trains its own DNN classifier as the surrogate model by observing the spectrum to detect the ongoing transmissions of the background emitter and generate adversarial attacks to fool the transmitter into misclassifying the channel as idle. This surrogate model may differ from the transmitter's classifier significantly because the adversary and the transmitter experience different channels from the background emitter and therefore their classifiers are trained with different distributions of inputs. This system model may represent a setting where the background emitter is a primary user, the transmitter is a secondary user, and the adversary is trying to fool the secondary user to transmit even though the channel is occupied by the primary user. We consider different topologies to investigate how different surrogate models that are trained by the adversary (depending on the differences in channel effects experienced by the adversary) affect the performance of the adversarial attack. The simulation results show that the surrogate models that are trained with different distributions of channel-induced inputs severely limit the attack performance and indicate that the transferability of adversarial attacks is neither readily available nor straightforward to achieve since surrogate models for wireless applications may significantly differ from the target model depending on channel effects.

\end{abstract}

\section{Introduction}\label{sec:Introduction}



Deep learning (DL) has been broadly applied to solve complex problems in various domains such as computer vision and speech recognition \cite{Goodfellow1} by learning from and adapting to rich data environments via deep neural networks (DNNs). On the other hand, it is well known that the DNNs are susceptible to the adversarial attacks (also known as evasion attacks) that can fool the DNNs into misclassification by adding small perturbations to the input data samples \cite{Szegedy1}. 
When the target DNN model is not readily available to the adversary, it can query the DL system (the target DNN model) with its own input data and use the returned labels to build a surrogate model (another DNN model). Then, the adversary uses this surrogate model to craft adversarial attacks against the target model by relying on transferability of adversarial attacks (namely, the attacks developed using a reliable surrogate model should also work against the target model with high success). In data domains such as computer vision, the adversary has typically the advantage of having (i) control of adding adversarial perturbations directly to the input data and (ii) direct access to training data for the surrogate model that have the same or similar distributions as the training data for the target model. 

These two properties do not necessarily hold in wireless communication systems, where DL has found rich applications such as spectrum sensing \cite{Davaslioglu2}, signal classification \cite{Dyspan2019}, and waveform design \cite{erpek1}. Recently, adversarial machine learning has gained attention to understand the emerging attack surface regarding wireless security. Those attacks include exploratory (inference) \cite{Shi2018AdDL4CogRaSec, Terpek}, adversarial (evasion) \cite{Larsson2, Kokalj2, Kokalj3, Flowers1, Bair1, Lin, Kim1, Kim2, Gunduz2, Kim5G, KimMultiple, IoT}, poisoning (causative) \cite{YiMilcom2018,Sagduyu1}, membership inference \cite{MIA}, Trojan \cite{Davaslioglu1}, and spoofing \cite{Shi2019generative, ShiGANSpoofing} attacks. 

Adversarial attacks  aim to fool a DNN into making misclassification errors by adding adversarial perturbations to the input data. In the wireless domain, it has been shown in \cite{Larsson2} that the DNN modulation classifier from \cite{Oshea3} is vulnerable to adversarial attacks generated using fast gradient method (FGM) when the adversarial perturbations are transmitted through an additive white Gaussian noise (AWGN) channel. Vulnerabilities of the modulation classifier in the presence of adversarial attacks have been studied in \cite{Kokalj2, Kokalj3, Flowers1, Bair1, Lin}. Adversarial attacks on the modulation classifier in realistic channel environments such as Rayleigh fading have been studied in \cite{Kim1,Kim2} by accounting for the adversary's lack of direct control on adversarial perturbations (namely, the adversarial perturbations need to go through complex channel effects before reaching the target classifier at the receiver). In addition to fooling the modulation classifier at one receiver, the requirement has been added in \cite{Gunduz2} to guarantee that another receiver can still recover the underlying message with high reliability. This concept has been also applied in \cite{Kim5G} to enable covert communications without being detected by signal classifiers. Adversarial attacks have been extended in \cite{KimMultiple} to the use of multiple antennas to craft and transmit perturbations. Adversarial attacks have been also considered against spectrum sensing in \cite{YiMilcom2018,Sagduyu1, IoT}, where the attack aims to manipulate the spectrum sensing data collected at the transmitter to make incorrect transmit (spectrum access) decisions using the surrogate DL model built by the adversary. 

\begin{figure}[t]
	\centerline{\includegraphics[width=\columnwidth]{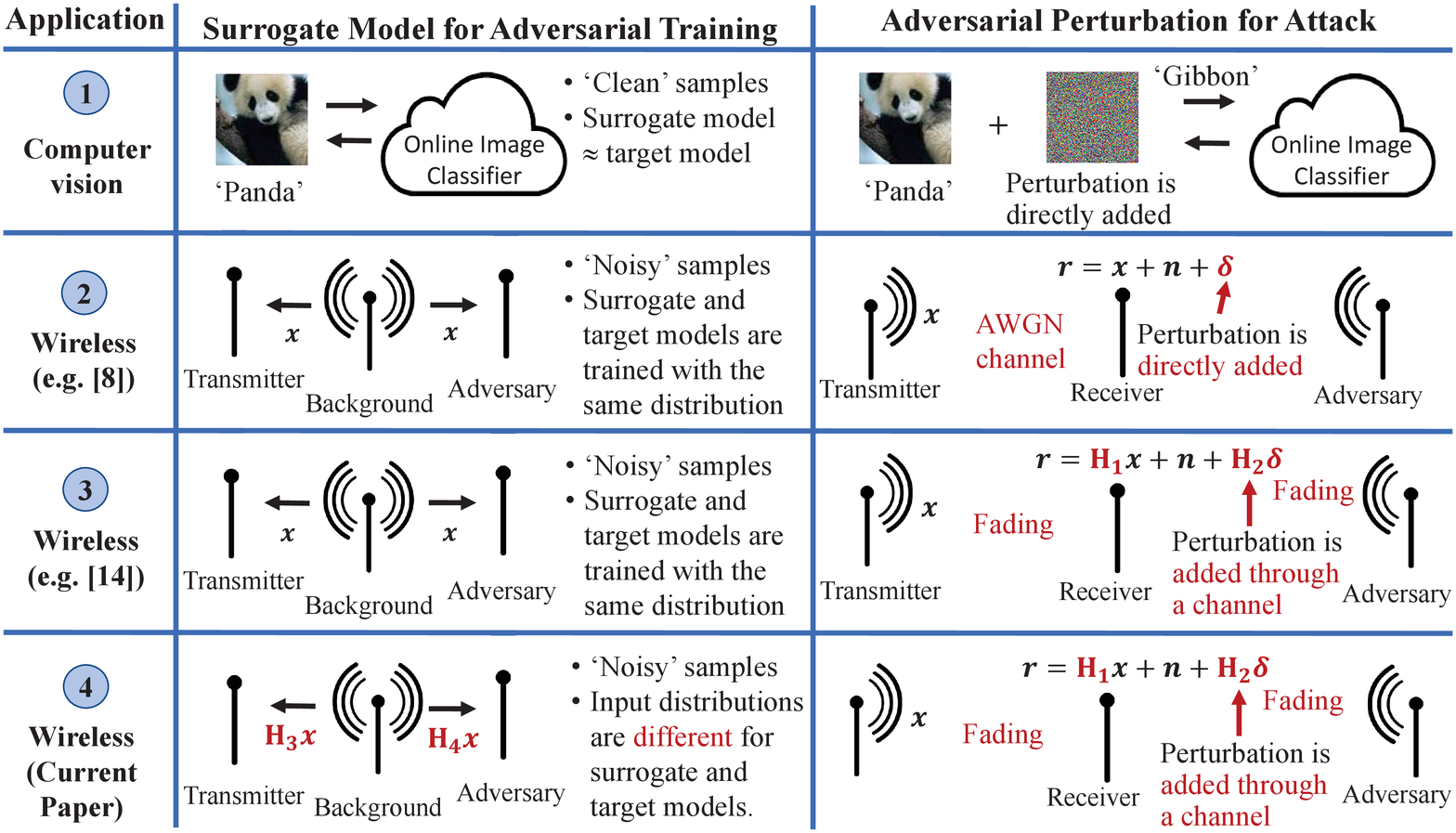}}
	\caption{Adversarial attacks from computer vision to wireless applications.}
	\label{taxonomy}
\end{figure}

For adversarial attacks on wireless signal classification, a common assumption has been made that the inputs to the target model and the surrogate model have the same distributions such that the adversarial attacks can be readily transferred from the surrogate model to the target model. However, practical wireless deployments require that the wireless adversary builds the surrogate model by observing the spectrum and collecting the training data over the air. Therefore, the surrogate model and the target model may not have the same input distributions due to the discrepancies in channels experienced by the adversary and the target receiver (e.g., due to different distances from the signal source that they are trying to sense). Fig.~\ref{taxonomy} summarizes the difference of adversarial attacks and surrogate models in different applications including computer vision and wireless communications.

In this paper, we investigate the channel effects on the surrogate model that is built by the adversary through over-the-air spectrum observations and used to craft adversarial attacks against a DNN-based wireless signal classifier. We consider a wireless communications system with a background emitter, a transmitter, and an adversary. The transmitter collects I/Q data and uses its DNN classifier to detect the ongoing transmission of the background emitter that is further used for channel access decisions. On the other hand, the adversary builds a surrogate model of the DNN classifier used at the transmitter. For that purpose, the adversary collects the I/Q data, namely signals received from the background emitter, by exploiting the broadcast nature of transmissions and obtains the labels by listening to the potential signals from the transmitter. If the transmitter classifies its received signal as `noise', then it transmits so the adversary obtains the label `noise'. If not, the label becomes `signal'. Note that there could be discrepancies between the labels of the target model and the surrogate model that we do not consider in this paper. Instead, we focus on the discrepancies between the input data samples for the target model and the surrogate model.  

Using the surrogate model, the adversary generates adversarial perturbations to fool the classifier at the transmitter from label `signal' to label `noise'. This setting may model a system where the background emitter is a primary user, the transmitter is a secondary user, and the adversary is trying to fool the secondary user to transmit even though the channel is used by the primary user. Compared to previous works, we make the realistic assumption that the adversarial attacks are generated using the surrogate model that is trained with a different distribution of the training dataset relative to the transmitter's target model since the signals received by the adversary and the transmitter are different due to channel discrepancies. Different topologies of the adversary are considered to investigate how the difference in the distribution of the training datasets affects the performance of the adversarial attack. Also, different approaches to select the transmit power at the adversary node using the surrogate model are considered. Finally, we relax the assumption that the adversary knows the exact input at the transmitter when determining the adversarial perturbation. Results show that the performance of adversarial attacks against a wireless signal classifier heavily relies on the reliability of surrogate model which depends on the difference of channels experienced by the adversary and the transmitter.
 
The rest of the paper is organized as follows. Section \ref{sec:SystemModel} provides the system model. Section \ref{sec:SDPA} describes how to generate adversarial attacks against wireless signal classifiers by using surrogate models. Section \ref{sec:topology} considers different topologies to evaluate the effect of surrogate models. Section \ref{sec:simulation} presents simulation results. Section \ref{sec:Conclusion} concludes the paper.

\section{System Model} \label{sec:SystemModel}

\begin{figure}[t]
	\centerline{\includegraphics[width=0.6\linewidth]{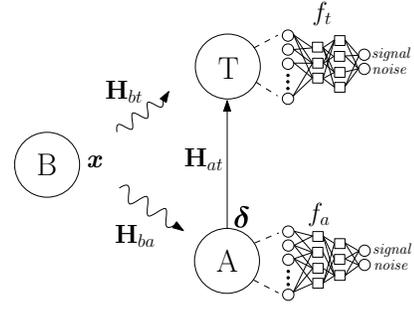}}
	\caption{System model. `B' stands for the background emitter, `T' stands for the transmitter, and `A' stands for the adversary.}
	\label{sys}
\end{figure}

We consider a wireless communication system that consists of a background emitter, a transmitter, and an adversary as shown in Fig. \ref{sys}. The background emitter transmits $k$ complex symbols, $\boldsymbol{x}\in \mathbb{C}^{k}$, and node $j$ (either the transmitter $t$ or the adversary $a$) receives
\begin{align}
\boldsymbol{r}_{bj} = \mathbf{H}_{bj}\boldsymbol{x} + \boldsymbol{n}_{bj}, \quad j \in \{t,a\},
\end{align}
where $\mathbf{H}_{bj} = \mbox{diag} \{h_{bj,1},\cdots, h_{bj,k}\}\in \mathbb{C}^{k\times k}$ and $\boldsymbol{n}_{bj}\in \mathbb{C}^{k}$ are the channel gains and complex Gaussian noise from the background emitter to node $j$, respectively.

Using $\boldsymbol{r}_{bt}$, the transmitter trains its DNN classifier, $f_{t}(.;\boldsymbol{\theta}_{t}): \mathcal{X} \rightarrow \mathbb{R}^{2}$, to determine the existence of ongoing background transmission to utilize the idle bands, where  $\boldsymbol{\theta}_{t}$ is the set of transmitter's DNN parameters and $\mathcal{X} \subset \mathbb{C}^{k}$. 
The input $\boldsymbol{r}_{bt}$ is assigned to the label $\hat{l}_{t}(\boldsymbol{r}_{bt},\boldsymbol{\theta}_{t}) = \arg \max_{q} f^{(q)}_{t}(\boldsymbol{r}_{bt},\boldsymbol{\theta}_{t})$, where $f_{t}^{(q)}(\boldsymbol{r}_{bt},\boldsymbol{\theta}_{t})$ is the output of classifier $f_{t}$ corresponding to the $q$th label.

Concurrently, the adversary tries to detect the background transmission based on $\boldsymbol{r}_{ba}$ using its own DNN classifier as the \emph{surrogate model}, $f_{a}(.;\boldsymbol{\theta}_{a}): \mathcal{X} \rightarrow \mathbb{R}^{2}$, where $\boldsymbol{\theta}_{a}$ is the set of the adversary's DNN parameters. The label corresponding to the input $\boldsymbol{r}_{ba}$ is defined as $\hat{l}_{a}(\boldsymbol{r}_{ba},\boldsymbol{\theta}_{a}) = \arg \max_{q} f^{(q)}_{a}(\boldsymbol{r}_{ba},\boldsymbol{\theta}_{a})$, where $f_{a}^{(q)}(\boldsymbol{r}_{ba},\boldsymbol{\theta}_{a})$ is the output of classifier $f_{a}$ corresponding to the $q$th label. If the label of the classifier is `signal', the adversary transmits a perturbation $\boldsymbol{\delta}\in \mathbb{C}^{k}$ to change the label of the DNN classifier at the transmitter to `noise'. Fooling the transmitter's classifier to change the label from `noise' to `signal' (by adding some interference) is easier than `signal' to `noise' (that requires careful signal cancellation), thus we only consider the more difficult case of changing the label from `signal' to `noise'.

If the background emitter transmits $\boldsymbol{x}$ and the adversary transmits $\boldsymbol{\delta}$, the received signal at the transmitter is
\begin{align}
&\boldsymbol{r}'_{bt}(\boldsymbol{\delta}) = \mathbf{H}_{bt}\boldsymbol{x}+\mathbf{H}_{at}\boldsymbol{\delta}+\boldsymbol{n}_{bt},
\end{align}
where $\mathbf{H}_{at}=\mbox{diag} \{h_{at,1},\cdot\cdot\cdot, h_{at,k}\}\in \mathbb{C}^{k\times k}$ is the channel gain from the adversary to the transmitter. The adversary generates the adversarial perturbation $\boldsymbol{\delta}$ to cause misclassification at the transmitter for the input $\boldsymbol{r}_{bt}$ while satisfying the power budget $P_{max}$ for a stealthy attack. Thus, the adversary determines $\boldsymbol{\delta}$ by solving the following optimization problem:
\begin{align} \label{eq:perturbation}
\argminA_{\boldsymbol{\delta}}& \quad ||\boldsymbol{\delta}||_2\nonumber\\
s.t. &\quad\hat{l}_{t}(\boldsymbol{r}_{bt},\boldsymbol{\theta}_{t}) \ne \hat{l}_{t}(\boldsymbol{r}'_{bt}(\boldsymbol{\delta}),\boldsymbol{\theta}_{t}) \nonumber \\ &\quad  ||\boldsymbol{\delta}||^{2}_{2} \le P_{max}. 
\end{align}

However, in reality, the adversary may not have information about the classifier at the transmitter, $f_{t}$, and it is not possible to check whether an attack generated at the adversary satisfies the constraint $\hat{l}_{t}(\boldsymbol{r}_{bt},\boldsymbol{\theta}_{t}) \ne \hat{l}_{t}(\boldsymbol{r}'_{bt}(\boldsymbol{\delta}),\boldsymbol{\theta}_{t})$, or not. Therefore, we change this constraint by using the DNN classifier at the adversary, ${f}_{a}$, which can be thought of as a \emph{surrogate model} for the transmitter's classifier. Now, the optimization problem to generate the adversarial perturbation at the adversary is
\begin{align} \label{eq:perturbation2}
\argminA_{\boldsymbol{\delta}}& \quad ||\boldsymbol{\delta}||_2\nonumber\\
s.t. &\quad\hat{l}_{a}(\boldsymbol{r}_{bt},\boldsymbol{\theta}_{a}) \ne \hat{l}_{a}(\boldsymbol{r}'_{bt}(\boldsymbol{\delta}),\boldsymbol{\theta}_{a}) \nonumber \\ &\quad  ||\boldsymbol{\delta}||^{2}_{2} \le P_{max}. 
\end{align}

As the non-linearity of the DNNs makes the solution $\boldsymbol{\delta}^*$ to (\ref{eq:perturbation2}) difficult to obtain, the adversarial perturbation is approximated. Fast gradient method (FGM) \cite{Kurakin1} is an efficient method to craft adversarial attacks by linearizing the loss function, $L_{j}(\boldsymbol{\theta},\boldsymbol{x},\boldsymbol{y})$, of the $j$th node's DNN classifier in a neighborhood of input $\boldsymbol{x}$, where $\boldsymbol{y}$ is the label vector, and uses linearized loss function for the optimization. In this paper, we consider a targeted attack, specifically the maximum received perturbation power (MRPP) attack in \cite{Kim1}, where the adversary's perturbation aims to decrease the loss function of data with label `noise' and cause a specific misclassification, from label `signal' to label `noise', at the transmitter in the presence of background transmission. 

\section{Adversarial Attack built upon the Surrogate Model of the Adversary}\label{sec:SDPA}
In this section, we introduce how to craft an adversarial perturbation using the surrogate model of the adversary, $f_a$, to flip the label of the transmitter's classifier. To do so, the adversary first tries to detect the background transmission based on $\boldsymbol{r}_{ba}$ using its own DNN classifier, $f_{a}$. Since the adversary tries to change the label from `signal' to `noise', the adversary generates an attack if the label $\hat{l}_{a}(\boldsymbol{r}_{ba},\boldsymbol{\theta}_{a}) = \arg \max_{q} f^{(q)}_{a}(\boldsymbol{r}_{ba},\boldsymbol{\theta}_{a})= $ `signal'. Then, the adversary crafts the adversarial perturbation using the MRPP attack. Note that we assume that the adversary knows the channel $\mathbf{H}_{at}$ and the exact input at the transmitter, $\boldsymbol{r}_{bt}$, when creating the adversarial perturbation. We relax the assumption regarding the knowledge of the exact input at the transmitter by using the input at the adversary, $\boldsymbol{r}_{ba}$, instead of $\boldsymbol{r}_{bt}$ later in the paper.   

Since the adversary has no information about the transmitter's classifier, it is hard for the adversary to determine the transmit power that is needed to change the label at the transmitter. Therefore, we consider two different ways to decide the transmit power at the adversary.

\subsection{Adversarial perturbation using the maximum power $P_{max}$}\label{IIIA}
Without knowing the transmitter's classifier, one simple option for the adversary is to use the maximum power for the adversarial perturbation. Therefore, the adversary simply transmits $\boldsymbol{\delta}^{*} = -P_{ max}\frac{\nabla_{\boldsymbol{x}}\mathbf{H}_{at}^{*}L_{a}(\boldsymbol{\theta}_{a},\boldsymbol{r}_{bt},\boldsymbol{y}^{\textit{target}})}{(||\nabla_{\boldsymbol{x}}\mathbf{H}^{*}_{at}L_{a}(\boldsymbol{\theta}_{a},\boldsymbol{r}_{bt},\boldsymbol{y}^{\textit{target}})||_{2})}$, where  $\boldsymbol{y}^{\textit{target}}$ is `noise'.

\subsection{Adversarial perturbation using the surrogate model of the adversary}\label{IIIB}
The transmit power at the adversary should be carefully determined to fool the transmitter's classifier. This has been done in \cite{Larsson1,Kim1} by using the transmitter's classifier. However, the adversary has no information about the transmitter's classifier. Therefore, the adversary determines the transmit power using its own classifier as a surrogate model for the transmitter's classifier. The details of how to generate the adversarial perturbation are presented in Algorithm \ref{alg1}.

\subsection{Adversarial perturbation using $r_{ba}$}
Previously, we assumed that the adversary knows the exact input at the transmitter, $\boldsymbol{r}_{bt}$. However, it is impractical to assume that the adversary knows $\boldsymbol{r}_{bt}$. Thus, the adversary uses the received signal at the adversary, $\boldsymbol{r}_{ba}$, instead of $\boldsymbol{r}_{bt}$ to craft the adversarial perturbation. The same procedure used in Section~\ref{IIIA} and Section~\ref{IIIB} can be used to generate the adversarial perturbation without the knowledge of the input at the transmitter by changing $\boldsymbol{r}_{bt}$ to $\boldsymbol{r}_{ba}$ in conjunction with using the surrogate model of the adversary. 

\begin{algorithm}[t]
	\DontPrintSemicolon
	\SetAlgoLined
	\label{alg1}
	Inputs: $\boldsymbol{r}_{bt}$, desired accuracy $\varepsilon_{acc}$ and  $P_{\textit{max}}$ \\
	Initialize: ${\varepsilon}\leftarrow {0}, \varepsilon_{\textit{max}} \leftarrow \sqrt{P_{\textit{max}}}, \varepsilon_{min} \leftarrow 0, \boldsymbol{y}^{\textit{target}} \leftarrow \text{`noise'}$ \\
	$\boldsymbol{\delta}_{norm} =\frac{\nabla_{\boldsymbol{x}}\mathbf{H}_{at}^{*}L_{a}(\boldsymbol{\theta}_{a},\boldsymbol{r}_{bt},\boldsymbol{y}^{\textit{target}})}{(||\nabla_{\boldsymbol{x}}\mathbf{H}_{at}^{*}L_{a}(\boldsymbol{\theta}_{a},\boldsymbol{r}_{bt},\boldsymbol{y}^{\textit{target}})||_{2})}$\\
	\If{$\hat{l}_{a}(\boldsymbol{r}_{ba})== \text{`noise'}$}{
	\While{$\varepsilon_{\textit{max}}-\varepsilon_{min} > \varepsilon_{acc}$}{
		$\varepsilon_{avg} \leftarrow (\varepsilon_{\textit{max}}+\varepsilon_{min})/2$\\
		$\boldsymbol{x}_{adv} \leftarrow \boldsymbol{r}_{bt} - \varepsilon_{avg}\mathbf{H}_{at}\boldsymbol{\delta}_{\textit{norm}}$\\
		\lIf{$\hat{l}_{a}(\boldsymbol{x}_{adv})==  \text{`noise'}$}{
			$\varepsilon_{min}\leftarrow \varepsilon_{avg}$}
		\lElse{$\varepsilon_{\textit{max}}\leftarrow \varepsilon_{avg}$}
	}}{$\varepsilon = \varepsilon_{\textit{max}}$, $\boldsymbol{\delta}^{*} = -\varepsilon\boldsymbol{\delta}_{norm} $}\\
	\caption{Adversarial perturbation using $\boldsymbol{r}_{ba}$}
\end{algorithm}

\section{Topologies to Characterize the Effects of Surrogate Models on Adversarial Attacks}\label{sec:topology}
We consider different topologies to investigate the effects of the surrogate models (used by the adversary at different locations) on the performance of adversarial attacks. We assume that the channel between node $i$ and node $j$ is Rayleigh fading with path-loss and shadowing, i.e., $h_{ij,k}=K(\frac{d_{0}}{d_{ij}})^{\gamma}\psi h^{ray}_{ij,k}$, where $ K = 1, d_{0}=1, \gamma = 2.7, \psi \sim\ $Lognormal$(0,8)$, ${h}^{ray}_{ij,k} \sim \mbox{Rayleigh}(0,1)$ and $d_{ij}$ is the distance between node $i$ and node $j$. Thus, if the location of the adversary changes with respect to the background emitter, the classifier at the adversary is trained with a different input distribution that changes with respect to the distance from the background emitter. Also, if the location of the adversary changes with respect to the transmitter, the usage of the power needed at the adversary changes, i.e., the adversary needs more power if the distance between the transmitter and adversary increases. To assess the performance of the classifier at the adversary when the location changes, we consider the following adversary locations while we fix the distance between the background emitter and the transmitter as $1$ in all topologies.

\subsection{Fixed distance between the background emitter and the adversary}
\begin{figure}[t]
	\centerline{\includegraphics[width=0.55\linewidth]{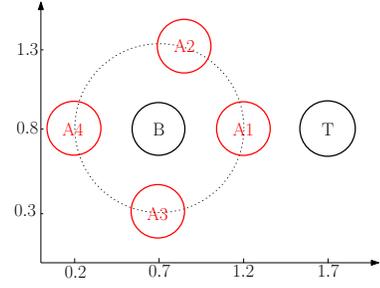}}
	\caption{Fix the distance from the background emitter `B' to the adversary (A1-A4 correspond to different locations of the adversary).}
	\label{fix pm}
\end{figure}

We first consider topologies where the distance between the background emitter and the adversary is fixed for all cases and the only difference among adversary locations is the distance to the transmitter as shown in Fig. \ref{fix pm}, i.e., the adversary uses the same surrogate model for these topologies. Specifically, the distance between the background emitter and the adversary, namely $d_{ba}$, is $0.5$ for adversary locations A1-A4 in Fig.~\ref{fix pm} and the distance between the transmitter and the adversary, namely $d_{ta}$, is $0.5, 1, \sqrt{1.25}, 1.5$ for adversary locations A1-A4 in Fig.~\ref{fix pm}, respectively. In this setting, the classifier of the adversary remains the same when its location changes, since the distributions of received signals from the background emitter are the same. The only difference among the adversary locations is the distance to the transmitter.    

\begin{figure}[t]
	\centerline{\includegraphics[width=0.55\linewidth]{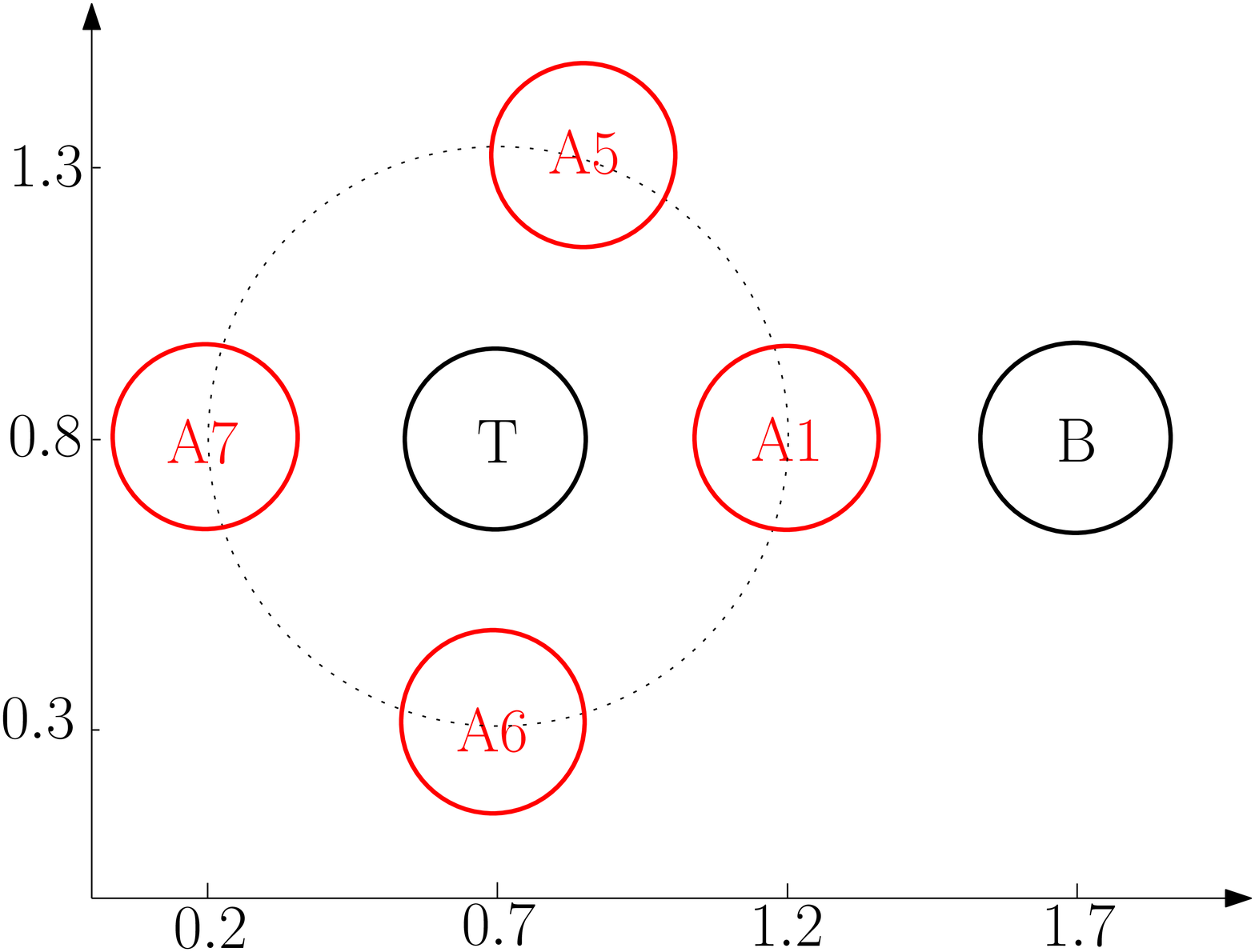}}
	\caption{Fixed distance between the transmitter `T' and the adversary (A1, A5-A7 correspond to different locations of the adversary).}
	\label{fix sm}
\end{figure}

\subsection{Fixed distance between the transmitter and the adversary}
Next, we consider the topology in Fig. \ref{fix sm}, where the distance between the adversary and the transmitter is fixed for all cases and the difference among the adversary locations is the distance to the background emitter, i.e., the surrogate models of the adversary are different, whereas the channel effects on the adversarial perturbations have the same distributions.
Specifically, $d_{ta}$ is $0.5$ for adversary locations A1, A5, A6, and A7 in Fig.~\ref{fix sm} and $d_{ba}$ is $0.5, 1, \sqrt{1.25}, 1.5$ for adversary locations A1, A5, A6, and A7 in Fig.~\ref{fix sm}, respectively. Since the locations of the adversary are different with respect to the background emitter, the classifiers of the adversary at different locations are trained with different distribution of inputs. Note that the classifier at the transmitter is trained with the same distribution of inputs as the adversary at position 5 (A5) and differently from other locations.  

\section{Performance Evaluation}\label{sec:simulation}
In this section, we investigate how different topologies described in Section \ref{sec:topology} and the corresponding channels affect the performance of the adversarial perturbation built upon the adversary's surrogate model. We assume that there is QPSK signal transmission in the background and the classifiers at the transmitter and the adversary are both a convolutional neural network (CNN), where the input to the CNN is of two dimensions (2,16) corresponding to 16 in-phase/quadrature (I/Q) data samples. We investigate the adversarial perturbation performance when the classifier at the adversary is the same as or different from the classifier at the transmitter. In all cases, the adversary trains a surrogate model that is different from the model of the transmitter's classifier. 

The default CNN structure used in the paper consists of a convolutional layer with kernel size $(1,3)$, a hidden layer with dropout rate $0.1$, ReLu activation function at convolutional and hidden layers and softmax activation function at the output layer that provides the label `signal' or `noise'. We apply the backpropagation algorithm with Adam optimizer to train the CNN using cross-entropy as the loss function. The CNN is implemented in Keras with TensorFlow backend. We start with the same classifier architecture for the transmitter and the adversary. Even in this case, each classifier is trained with a different input data distribution due to the different channel between the background emitter and each node. We use the perturbation-to-noise ratio (PNR) metric that is also used in \cite{Larsson1,Kim1} to represent the relative perturbation power with respect to the noise. Note that as the PNR increases, it becomes easier to detect the adversary's transmission.

\begin{figure}[t]
	\centerline{\includegraphics[width=0.8\linewidth]{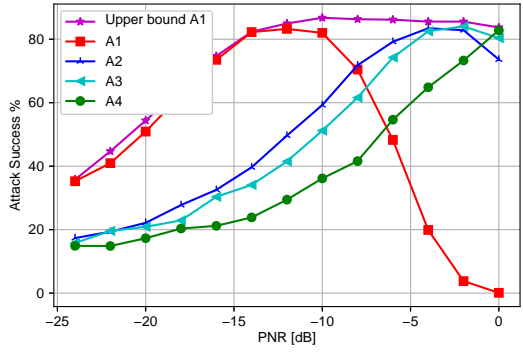}}
	\caption{Attack performance of the adversary using maximum power when $d_{ba}$ is fixed.}
	\label{sim fix pm}
\end{figure}
\begin{figure}[t]
	\centerline{\includegraphics[width=0.8\linewidth]{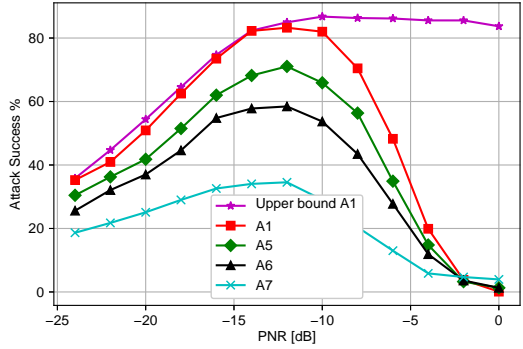}}
	\caption{Attack performance of the adversary using maximum power when $d_{ta}$ is fixed.}
	\label{sim fix sm}
\end{figure}

In Fig. \ref{sim fix pm}, we first consider the topology depicted in Fig.\ref{fix pm}, where the distance between the background emitter and each adversary location is the same. We use the same classifier for the adversary at all locations in this topology since the input distributions for the adversary at all these locations are the same and the only difference is its distance to the transmitter. The case where the classifier of the adversary is the same as the transmitter's classifier is considered as an upper bound where the power is determined using Algorithm 1. As the distance between the adversary and the transmitter increases, the peak of the curve shifts to the right meaning that it needs more power to fool the classifier at the transmitter. Moreover, the attack success first increases as the PNR increases and then decreases after exceeding some PNR value. This is because the adversary transmits the adversarial perturbation with the maximum power that fails to fool the classifier at the transmitter into classifying the received signal as `noise' when too much power is used at the adversary. Therefore, the adversary should select the right amount of power to transmit the adversarial perturbation. When the surrogate model at the adversary is the same as the transmitter's classifier, the attack success increases up to some PNR value and then saturates after exceeding it.

Next, we consider a different topology, where the distance between the transmitter and each adversary location is the same as shown in Fig. \ref{fix sm}. Since the distance between the background emitter and each adversary location is different, the distributions of the received signal at different adversary locations are different. Thus, we train the classifiers for the adversary at different locations differently and use these classifiers in the simulations. In Fig. \ref{sim fix sm}, as the distance between the background emitter and the adversary increases, the peak of the attack success decreases. In addition, even though the adversary at location 5 is trained with the same distribution as the transmitter's classifier, the classifier of the adversary at location 1 that is closer to the background emitter performs better. This observation suggests that the adversary should be located closer to the background emitter when the distance to the transmitter is fixed to increase the attack performance. 

\begin{figure}[t]
	\centerline{\includegraphics[width=0.8\linewidth]{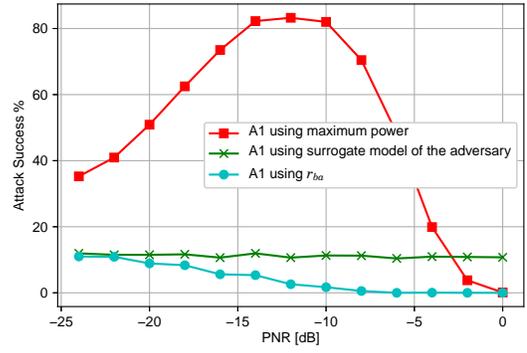}}
	\caption{Attack performance of adversary using maximum power, surrogate model of the adversary, or $\boldsymbol{r}_{ba}$.}
	\label{sim diff method}
\end{figure}

\begin{figure}[t]
	\centerline{\includegraphics[width=0.8\linewidth]{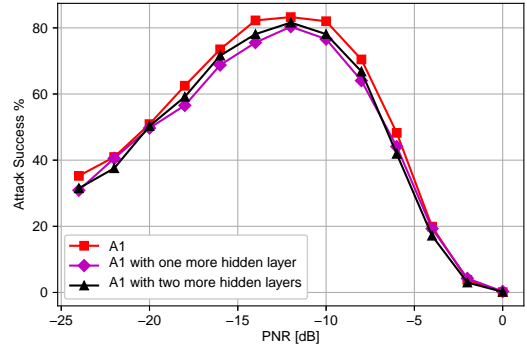}}
	\caption{Attack performance of adversary with different classifier architectures.}
	\label{sim diff classifier}
\end{figure}

The performance of the adversarial attack using maximum  power, the surrogate model, or $\boldsymbol{r}_{ba}$ is compared in Fig. \ref{sim diff method}. Even though the adversary precisely decides the transmit power based on its own classifier, this attack performs poorly compared to the case when the adversary uses maximum power. When we relax the assumption of knowing $\boldsymbol{r}_{bt}$ and instead use $\boldsymbol{r}_{ba}$ to generate the adversarial attack at the adversary, the overall performance drops compared to other methods and decreases when the PNR increases. 

Finally, we apply a DNN architecture for the surrogate model at the adversary that is different from the transmitter's classifier. The results are shown in Fig. \ref{sim diff classifier}. Although the number of hidden layers increases compared to the CNN with one hidden layer, the impact on the attack success is negligible. The transferability \cite{Transfer1} property holds since the performance has not changed although the architecture changes at the adversary. However, the surrogate models may be significantly different from the target model as the channels experienced by the transmitter and the adversary may differ. There may be other channel differences, such as multipath, intersymbol interference (ISI), and mobility (Doppler), present in practice. These differences may further increase the difference between the target model and the surrogate model at the adversary, and reduce the attack success. Therefore, the transferability argument is not readily applicable in practical wireless applications of over-the-air adversarial attacks.

\section{Conclusion} \label{sec:Conclusion}
We considered a wireless communication system where an adversary transmits a perturbation signal to fool the DNN classifier at a transmitter into classifying the ongoing background transmission as noise. The adversary trains its surrogate model by observing the spectrum and uses this model to design the adversarial attack. Through different topologies, we showed how the adversary's location significantly affects the attack performance as the surrogate model may differ from the target model due to channel discrepancies. In particular, the attack success against the transmitter drops when the adversary moves away from the background emitter (and the surrogate model becomes less reliable) although the adversary does not necessarily move closer to the transmitter. 


\bibliographystyle{IEEEtran}
\bibliography{lib}

\end{document}